\documentclass[twocolumn,a4paper,nofootinbib,showpacs,aps,floatfix]{revtex4}
\usepackage{graphicx}
\usepackage{bm}
\usepackage{amsmath}
\newcommand{\ignore}[1]{}

\def\beq{\begin{equation}}
\def\eeq{\end{equation}}
\def\la{\langle}
\def\ra{\rangle}
\begin{document}
\title{Long time deviation from exponential decay: 
non-integral power laws.} 
\author{J. Martorell}
\email{martorell@ecm.ub.es}
\affiliation{Departament
  d'Estructura i Constituents de la Materia, Facultat F\'{\i}sica,\\
   University of Barcelona, Barcelona 08028, Spain}
\author{J. G. Muga}
\email{jg.muga@ehu.es}
\affiliation{
Departamento de Qu\'\i mica-F\'\i sica, UPV-EHU, Apdo 644,   
  48080 Bilbao, Spain}
\author{D. W. L. Sprung}
\affiliation{
Department of Physics and Astronomy, McMaster University\\
  Hamilton, Ontario L8S 4M1 Canada}
%
\begin{abstract}
Quantal systems are predicted to show a change-over from exponential 
decay to power law decay at very long times. Although most 
theoretical studies predict integer power-law exponents, recent 
measurements by Rothe {\it et al}. of decay luminescence of organic 
molecules in solution \{Phys. Rev. Lett. {\bf 96} (2006) 163601\} 
found non-integer exponents in most cases. We propose a physical 
mechanism, within the realm of scattering from potentials with long 
tails, which produces a continuous range of power law exponents. In 
the tractable case of the repulsive inverse square potential, we 
demonstrate a simple relation between the strength of the long range 
tail and the power law exponent. This system is amenable to 
experimental scrutiny.  
\end{abstract}
\maketitle
%
%
\section{Introduction}

Rothe {\it et al}. \cite{RHM06} have recently presented experimental 
evidence for ``Violation of the Exponential Decay-Law at Long 
Times''. Deviations from exponential decay of an 
unstable quantum system had been predicted long ago as a result of the 
semiboundedness of the energy spectrum \cite{Khalfin}. However, the 
transition into the non-exponential regime at long times occurs in 
most decaying systems after many lifetimes, where the minute survival 
probability had previously frustrated verification. In 
addition, it had been argued that interaction with the 
environment and/or the measurement process would suppress the long 
time deviation \cite{FGR78}
\footnote{In fact the perturbing effect of measurement depends on the 
particular process. It has been shown to be negligible for distant 
detectors in solvable models \cite{DMG06}.}.    
Rothe {\it et al}. measured the decay of luminescence of various 
dissolved organic molecules after pulsed laser excitation. The 
results clearly show a range of times over which the decay of the 
survival probability $P(t)$ is exponential, 
followed by another where it is algebraic, $P(t)\sim t^{-\mu}$. 

One puzzling aspect of the experimental results of Rothe {\it et al}.  
\cite{RHM06} is that their inferred algebraic exponents $\mu$  are 
patently non-integral, except in one case (PtOEP, frozen solution), 
for which $\mu= 4$. Other values range from $\mu= 2.08$ to 
$4.07$. These non-integer exponents are in sharp contrast to
theoretical studies of long-time deviation from exponential decay 
which predict or postulate integral exponents for the algebraic decay 
\cite{FGR78,Hellund53,JS61,Winter61,Newton61,GW64,Ros65,KM76,Kn77,OK92,MDS95,MWS96,GC01,Dicus02,DN04,Miy04,Lon06}. 

Reacting to this apparent contradiction, we explore physical 
mechanisms which could lead to long-time deviations from exponential 
decay with arbitrary, non-integer power law exponents. On the 
mathematical side, an ``explanation'' of an arbitrary power-law 
exponent follows from the expression for the survival amplitude  in 
terms of an energy density $\omega(E)$. As defined in Section  II,  
\beq 
A(t)= \int_0^\infty dE\, \omega(E)\ e^{-iEt/\hbar}, 
\label{eq:z1}
\eeq
where we have assumed no bound states and an energy threshold at 
$E_m=0$.  The large-time asymptotics depend on the behaviour of 
$\omega(E)$ near the origin \cite{Kn77}. In fact, under mild 
conditions on $\omega(E)$ as a function of the complex variable $E$, 
$A(t)\sim t^{-(\nu+1)}$ as $t\to\infty$ if $\omega(E)\sim E^\nu$ as 
$E\to 0$. This implies $P(t)\equiv|A(t)|^2\sim t^{-2(\nu+1)}$. From 
this perspective, non-integer values of the exponent $\mu= 2(\nu+1)$ 
are the consequence of an appropriate non-integer $\nu$ in the 
threshold density of states. 

By itself, this formal argument is not satisfying since it tells us 
nothing about the physical mechanism behind the threshold behaviour 
of $\omega(E)$.  We would like to learn, from the properties of the 
Hamiltonian or possibly of the initial state, the origin of such 
exponents, or, equivalently, the threshold behaviour of the energy 
distribution.  At this point we refrain from attempting a very 
general approach valid for the plethora of unstable systems with 
exponential decay at intermediate times. In particular, the process 
of excitonic decay in dissolved organic molecules is poorly 
understood, and present models explain only qualitatively the 
processes that lead to their emission spectra 
\cite{BBS02,Bar02,Bar04}. Hence we take refuge in a  
physical domain where full or sufficiently accurate analytical and 
numerical methods are available, namely, the decay of a wavepacket in 
a spherically symmetric potential well, in a given partial wave.    
This is done with the hope that the lessons learned here may be 
useful elsewhere, although there is no guarantee that our results can 
be extended to such diverse fields as auto-ionizing states, 
photo-ionization, particle physics or spontaneous photon emission, because 
of the particular constructions of their distributions $\omega(E)$.

One of the advantages of the decay of unstable states in 1D and 
spherically symmetric 3D potential models is that it is a well 
studied subject, reviewed by Fonda, Ghirardi and Rimini some time ago 
\cite{FGR78}. The conditions leading to integer exponents are well 
understood. If the  initial unstable state has angular momentum 
$\ell$ and the potential is ``well behaved'' at short and large 
distances, then the algebraic exponent is generally $\mu= 2\ell+3$. 
In particular, for the most studied case of $s$-waves, $\mu=3$; the 
$\ell=0$ contribution would be in any case the dominant one at very 
large times since it is the slowest decaying term. This algebraic 
decay law  also applies in  1D decaying systems \cite{MDS95}. The 
derivation of these integral power laws will facilitate 
identification of the complementary cases in which the exponents may 
be non-integers, so we shall devote Section II to a review of the 
basic elements of potential scattering theory relevant for our 
purposes. (This presentation differs from \cite{FGR78} in some 
respects, but the results are equivalent.) We shall conclude that 
potentials with long tails are natural candidates for providing 
non-integral decay exponents. The case of inverse square potentials, 
$V(r) =2m \beta(\beta+1)/(\hbar^2 r^2)$ for $r > r_d$, and $\beta > 0$ 
(repulsive tail), is particularly suitable for theoretical treatment and 
provides as a first approximation a very simple relation between the 
strength  of the interaction and the exponent, $\mu = 2\beta+3$ 
(The attractive tail case ($-1/2< \beta <0$) will also be studied but 
the decay law is more involved.) The predicted algebraic exponent 
varies continuously with $\beta$, with $\mu = 3$ when $\beta=0$ 
corresponding to a short range potential.  This is shown in detail in 
Section III, for a  potential model with inverse square tail in which 
$\omega(E)$ is obtained analytically, and  the conditions for the 
validity of the simple prediction for the algebraic exponent can be 
easily studied.  In the final discussion we comment on the 
experimental feasibility of inverse square interactions. 

\section{Power laws for long time decay in potential scattering}
We consider a system that 
is initially in a normalized non-stationary state $|\Psi_0\ra$. The 
survival amplitude of that state is defined as the overlap of the 
initial state with the state at time $t$, and is the 
expectation value of the time evolution operator  
\begin{equation}                       
A(t) =  \la\Psi_0| {\rm exp} (-iHt/\hbar)|\Psi_0\ra,  
\label{eq:z2}
\end{equation}
$H$ being the Hamiltonian. The survival probability, sometimes 
called ``non-decay probability''\cite{FGR78}, is 
\begin{equation}                       
P(t) = |A(t)|^2. 
\label{eq:z3}
\end{equation}
This is the function whose asymptotic time behaviour we will study. 
The stationary states of the Hamiltonian, $H |\Phi_{E,\lambda}\ra = E 
|\Phi_{E,\lambda}\ra$, where by $\lambda$ we indicate any other quantum 
numbers characterizing states of energy $E$, determine a basis. 
In our models we have only a continuous spectrum, so the range of 
energies will be $E_m < E < \infty$, and the origin of energies 
is taken at threshold $E_m =0$. The completeness relation is then 
\begin{equation}                       
\sum_{\lambda} \int_0^{\infty} \ dE \ 
|\Phi_{E,\lambda}\ra\la\Phi_{E,\lambda}| = {\bf 1}, 
\label{eq:z4}
\end{equation}
which gives  
\begin{eqnarray}                       
A(t) &=& \int_{0}^{\infty}  dE  \ \omega(E) \ e^{-iEt/\hbar}
\nonumber \\
\omega(E) &=& \sum_{\lambda}|\la\Phi_{E,\lambda}|\Psi_0\ra|^2.
\label{eq:z5}
\end{eqnarray}
We will call $\omega(E)$ the ``energy density'' or ``energy 
distribution'' of the initial state. To construct  $\omega(E)$, we 
therefore need the continuum solutions denoted $|\Phi_{E, \lambda}\ra$. 
For a spherically symmetric potential, the Schr\"odinger equation has 
separable solutions $\Phi({\vec r}) = u_{\ell}(r) Y_\ell^m
(\Omega)/r$. For each partial wave 
\begin{eqnarray}                       
{{d^2 u_\ell}\over {dr^2}} -{{\ell(\ell+1)}\over r^2} u_\ell 
+ [k^2-v(r)] u_\ell  &=&  0,  
\label{eq:z6}
\end{eqnarray}
where $v(r) = (2m/\hbar^2) V(r)$ and $k^2 = (2m/\hbar^2) E$. 
For convenience, in the following we work with solutions
$w_\ell(k,r)$ normalized as 
\beq                                   
\int_0^\infty dr\,w_{\ell}(k',r)^*w_{\ell}(k,r)=
\delta(k'-k)
\label{eq:z7}
\eeq
and obeying the boundary condition 
\beq
\lim_{r\to\infty} w_\ell(k,r)=(2/\pi)^{1/2}\sin(kr-\pi\ell/2+\delta_\ell),
\label{eq:z8}
\eeq
where $\delta_\ell$ is the phase shift for the partial wave $\ell$. 
These solutions are related to the 
regular solutions $\hat{\phi}_\ell$ (defined by their behaviour 
as Riccati-Bessel functions $\hat{j}_\ell$, $\hat{\phi}_\ell(r)\sim
\hat{j}_\ell(kr)$ when $r\to 0$) by 
\beq
w_\ell=(2/\pi)^{1/2} \frac{\hat{\phi}_\ell}{|f_l(k)|},  
\label{eq:z9}
\eeq
where $f_l(k)$ is the Jost function. Among the different definitions  
we use the one by Taylor \cite{Taylor} so that
\beq                                  
f_\ell(k)=1+\frac{1}{k}\int_0^\infty dr \hat{h}_\ell^+(kr)v(r) 
\hat{\phi}_\ell(k,r),
\label{eq:z10}
\eeq
see the Appendix A for a minimal account of Riccati-Bessel
functions. 

We now sketch the arguments that justify integral power laws 
for the long time deviation from exponential decay in  
scattering from ``well-behaved'' potentials, defined as those falling off 
faster than $r^{-3}$ at infinity and by being less singular than 
$r^{-3/2}$ at the origin \cite{Taylor}.  We shall assume for 
simplicity that the initial non-stationary state is in a particular 
partial wave $\ell$ without bound states, and localized, 
\beq                                   
u_i(r) =0\; \ {\rm{for}}\; r>r_a. 
\label{eq:z11}
\eeq
Then the survival amplitude takes the form
\begin{eqnarray}                       
A(t) &=&\frac{2m}{\pi \hbar^2}\int_0^\infty dE \frac{1}{k}
\left|\frac{\la u_i |\hat{\phi}_{\ell}\ra}{f_\ell(k)}\right|^2e^{-iEt/\hbar}. 
\label{eq:z12}
\end{eqnarray}
The function (\ref{eq:z10})   
for $\ell>0$ vanishes at $k=0$ if and only if there is a zero energy 
bound state whereas for $\ell=0$ such an occurrence  represents a 
zero-energy resonance \cite{Taylor}. We disregard these exceptional 
possibilities and concentrate on the generic case ${f}_\ell(0) \neq 0$. 

The main properties of concern for the long-time behaviour of the 
survival probability are analyticity of $\hat{\phi}_\ell$ in the 
complex $p$-plane, its asymptotic behaviour near the origin (the same 
as the Ricatti-Bessel function) and at infinity \cite{Taylor}, 
\beq                                   
|\hat{\phi}_{\ell}(k,r)|\leq 
\gamma_\ell\left(\frac{|kr|}{1+|kr|}\right)^{\ell+1}e^{|{\rm Im} kr|}, 
 \label{eq:z13}
\eeq
where $\gamma_\ell$ is some constant, 
and  the behaviour of the Jost function at $k=0$ and at infinity. 
Eq. (\ref{eq:z10}) can be continued analytically in the upper $k$-plane 
(zeroes on the positive imaginary axis would represent bound 
states), and tends to one uniformly as $|k|\to\infty$ in Im$( k) \geq 0$. 

Now we deform the energy contour from the positive axis in 
Eq. (\ref{eq:z12}) to one directed upwards along the lower imaginary 
axis on the first energy sheet. A long closing arc at infinity 
does not contribute because the exponential growth of $\hat{\phi}_{\ell}$ 
implied in Eq. (\ref{eq:z13}) is limited by the localization 
(\ref{eq:z11}) and thus compensated by the decaying time 
exponent. The behaviour of the integrand near threshold is thus, 
from Eq. (\ref{eq:z13}) and the zero energy asymptotics of  the Jost 
function, of order $E^{\ell+1/2}$, so the long-time asymptotic of the 
survival amplitude is, according to Watson's lemma, $A\sim t^{-
(\ell+3/2)}$, and therefore 
\beq                                    
P(t)\sim t^{-(2\ell+3)}.
\label{eq:z14}
\eeq   
This is easily generalized to cases with a zero energy resonance or bound 
states, since the zero energy behaviour of the Jost function is known 
in these cases \cite{Taylor}, and again provides integer exponents.    
Another type of exception arises from the possible cancellation of the 
threshold  dependence in Eq. (\ref{eq:z13}) by integration over the 
coordinate in $\la \hat{\phi}|u_i\ra$, by a suitable choice of 
$u_i(r)$.    In the context of 1D potential models, 
Miyamoto \cite{Miy04} has studied the long time behaviour of 
wavepackets scattered by a  finite range potential. He showed that by 
careful adjustment of the low momentum components of the initial wave 
packet, initially located outside the interaction region, an exponent 
5 for the decay power law may occur, instead of the generic value 3. 
As presented, such state manipulation produces only odd integer values. 

Comparing the result (\ref{eq:z14}) with the dependence on $\ell$ 
of the centrifugal barrier, it is tempting to speculate that a $v(r)= 
\beta(\beta+1)/ r^2$ potential tail, where $\beta > 0$, 
may produce long-time decay power laws of the form $t^{-(2\beta+3)}$ 
for $\ell=0$, based on the fact that the long tail plays for large 
$r$ the role of a centrifugal term with a non-integer, effective 
$\ell$.  That this is indeed the case, will 
be demonstrated in a solvable model below. A proof of generality of
this result is however far 
from trivial, since the properties of the functions used to arrive at 
Eq. (\ref{eq:z14}) depend on fast fall-off conditions which are 
not satisfied for an inverse square tail. Instead of using $\ell=0$, 
we could alternatively assume from the start a non-integer, effective 
$\ell=\beta$ and its corresponding ``partial wave'' stationary 
equation to extend the usual results, but this requires a 
redefinition of the potential $V(r)$ in the internal region to 
compensate exactly for the unphysical inverse square term. The 
consequence is an inverse square divergence in $V(r)$ at $r=0$ which, 
again, makes the formal treatment non-standard.  We shall therefore 
be content with a demonstration, via exactly soluble models, that the 
proposed long-time decay behaviour with non-integral exponents does 
indeed occur, and leave a more general and abstract theory for a 
further publication.


%
%
%
\section{Decay of an unstable state in potential models}

\subsection{General results}
Consider a point particle of mass $m$ in 3D subject to a central 
force. First we will derive some general results assuming only that beyond a 
certain radius, $r=r_d$, the potential is inversely proportional to 
the square of the distance. 
We call this a ``boundary condition model''. 
For convenience, we write 
\begin{equation}                           
V(r) = {{2m}\over \hbar^2} \ {\beta(\beta+1)\over r^2} \quad , \quad r> r_d. 
\label{eq:z15}
\end{equation}
\begin{figure}[htb]                          
\includegraphics[height=6cm]{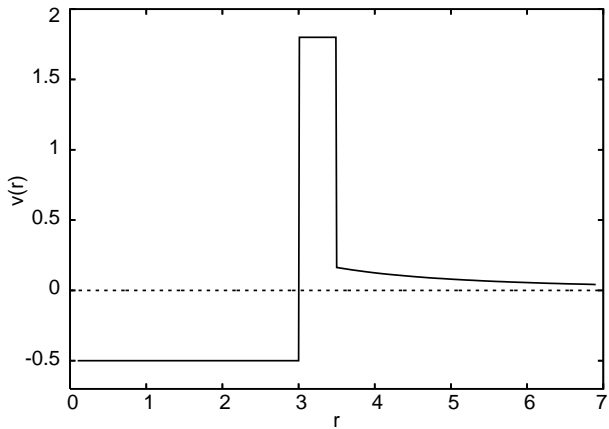}
\caption[]{Potential $V(r)$ for one illustrative example: 
$\hbar=2m=1$, $v_0= 0.5, v_b = 1.8, r_a=3.0, r_d=3.4$ and $\beta= 
1.$} 
 \label{figbar1}
\end{figure} 

The inner part is also assumed to be free of singularities, even 
at the origin, so that the existence of regular solutions of the 
Schr\"odinger equation is guaranteed. To have a vanishing survival 
probability when $ t \to \infty$, we choose a potential with no bound 
states.  Later, in order to have explicit analytic expressions for 
the properties of $\omega(E)$ and the asymptotic survival 
probability, we will use the more detailed ``WB'' (well-barrier) 
model. In that case, the inner part of $V(r)$ will be a square well, depth 
$- V_0 <0$ and extending from $0$ to $r_a$, followed by a square 
barrier of height $V_b >0$ located between $r_a$ and $r_d$. An example is
shown in figure \ref{figbar1}. With 
suitable parameters for the well and barrier this model has well 
defined quasibound states, which lead to the exponentially decaying 
part of $P(t)$.

As announced in the Introduction, we will show that the value of 
$\beta$ determines the exponent of the algebraic decay at long times. 
In  our examples $\beta > -1/2$ will be either positive 
(repulsive), or negative but such that the attraction does not 
support bound state solutions. Since 
$V(r) \to 0$ when $r \to \infty$,  the energy threshold in the model 
is indeed at $E_m=0$.

For simplicity we take the initial state to have $\ell=0$, and drop 
the subindex $\ell$ in the following.
It is also convenient to define a different regular solution 
$\phi(k;r)=\hat{\phi}(k,r)/k$  
that satisfies the boundary conditions 
\begin{equation}                       
\phi(k;0)=0 \quad , \quad \phi'(k;0)= 1
\label{eq:z16}
\end{equation}
and, as shown in \cite{GW64,MM65}, 
provided that 
the potential goes to zero faster than $1/r$ when $r\to \infty$, 
its asymptotic form is, see Eqs. (\ref{eq:z8},\ref{eq:z9}),  
\begin{equation}                       
\phi(k,r) \simeq
{{|f(k)|}\over k} \sin(kr+\delta).
\label{eq:z17}
\end{equation}                          
where $f(k) \equiv |f(k)| e^{-i\delta} = f^*(-k)$ on the real $k$-axis.  
In addition, to be consistent with Eq. (\ref{eq:z4}), we 
define solutions normalized with respect to energy, $w_E(r)$, 
\begin{equation}                        
w_E(r) = \sqrt{m\over {\hbar^2 k}} w(k;r) =\sqrt{{2m k}\over{\pi 
\hbar^2}} {1 \over{|f(k)|}}  \phi(k;r),  
\label{eq:z18}
\end{equation}
such that 
\begin{equation}                        
\la w_E|w_{E'}\ra=\delta(E -E') = {m\over {\hbar^2k} } \delta(k-k').
\label{eq:z19}
\end{equation}
Finally, we assume that the chosen potential allows a regular 
solution at zero energy, denoted $\phi_0(r) = \phi(0;r)$. 

For the potential in Eq. (\ref{eq:z15}), the regular solution for 
$r>r_d$ can be written as a linear combination of Riccati-Bessel functions, 
\begin{equation}                        
\phi(k,r) =  a_{III}\ \hat{j}_{\beta}(kr) + b_{III}\ \hat{n}_{\beta} (kr). 
\label{eq:z20}
\end{equation}
(The reason for the subindex $III$ is the 
consideration of two spatial regions for the inner part of the potential in 
the following subsection.)

For repulsive potentials $\beta >0$. We will also consider weakly 
attractive potentials with  $\beta > -1/2$ so that the order of the 
related cylinder functions $\nu=\beta+1/2$ remains positive; see 
Appendix A. For a detailed and updated discussion of the problems 
arising when the $1/r^2$ potential becomes too strongly attractive, 
$\beta<-1/2$, see reference \cite{CEF00}. The constants $a_{III},\, 
b_{III}$, can be written in terms of the regular solution and its 
derivative at $r_d$ by matching the inner and outer solutions at this 
boundary, 
\begin{eqnarray}                       
\phi(k,r_{d})\!&=&\!a_{III}\ \hat{j}_{\beta}(kr_d) + b_{III}\ \hat{n}_{\beta} 
(kr_d)\,
\nonumber \\
{{\partial \phi(k,r)}\over {\partial r}}\bigg|_{r=r_d}\!&=&\!\! 
k\!\left[ a_{III}\ \hat{j}'_{\beta}(kr_d)+  b_{III}\ 
\hat{n}'_{\beta}(kr_d)\right]\!\!, 
\label{eq:z21}
\end{eqnarray}
where the prime means derivative with respect to the argument of the Bessel
function, $x= kr$. 

Using their asymptotic forms and comparing to Eq. (\ref{eq:z17}),  
\begin{equation}                       
|f(k)| = k \sqrt{a_{III}^2 + b_{III}^2},
\label{eq:z22}
\end{equation}
with
\begin{eqnarray}                       
{\cal C}^2 &\equiv&  a_{III}^2+  b_{III}^2 \nonumber \\ 
&=& [({\hat n}'_{\beta})^2+ ({\hat j}'_{\beta})^2] \phi(k,r_d)^2 
+ {1\over k^2} \phi'(k,r_d)^2 (\hat{n}_{\beta}^2+\hat{j}_{\beta}^2) \nonumber \\
&-& {2\over k} \phi(k,r_d) \phi'(k,r_d) (\hat{n}_{\beta} \hat{n}'_{\beta} + 
\hat{j}_{\beta} \hat{j}'_{\beta}), 
 \label{eq:z23}
\end{eqnarray}
where, to simplify notation, $\phi'(k,r_d) = (\partial 
\phi(r,k)/\partial r)_{r_d}$,  and the argument of the  Bessel 
functions is $x_d = k r_d$. In subsection C we will need the 
asymptotic forms when $k \to 0$. In this limit,  $\phi(k,r_d) \to 
\phi_0(r_d) $, which will always be finite and non vanishing unless 
we have made a rather unfortunate choice for the point $r_d$. The 
value of the  derivative will also be finite. Eqs. (\ref{eq:z22}) and 
(\ref{eq:z23}) then show that in this limit the explicit $k$ 
dependence of the Jost function is determined by that of the 
Ricatti-Bessel functions: Eq. (\ref{eq:z34}) gives an example of this. 

Since our initial state has $\ell=0$, the sum over 
$\lambda$  in Eq. (\ref{eq:z5}) may be dropped, and writing 
the radial part of the initial state as  $|u_i\ra$, the energy 
density is 
\begin{equation}                      
\omega(E) = |\la w_E|u_i\ra|^2. 
\label{eq:z24}
\end{equation}
Once $|u_i\ra$ is chosen we can expand it in terms of 
the various solutions of the Schr\"odinger 
equation just described. We choose an initial state whose 
wavefunction is non-vanishing only when $r < r_a$,  
\begin{equation}                     
u_i(r) = \sqrt{2\over r_a}\ \sin k_a r \ \Theta(r_a-r),
\label{eq:z25}
\end{equation}
with $k_a= n_a \pi/r_a$ and $n_a$ integer. In this paper we 
take $n_a=1$. Then $|u_i\ra$  coincides with the 
ground state of the inner well in the limit of infinite barrier 
height. This initial state is widely used in simulations of 
unstable systems. Its advantage is its simple analytic form.
At intermediate times only the lowest quasibound state contributes 
significantly, which leads to exponential decay of $P(t)$. For this 
choice of $|u_i\ra$ 
\begin{equation}                     
\la w_E|u_i\ra = {2\over \hbar} \sqrt{{m k}\over{\pi r_a}} 
{{1}\over{|f(k)|}} \int_0^{r_a} \phi(k;r) \sin k_a r \ dr, 
 \label{eq:z26}
\end{equation}
which, combined with Eq. (\ref{eq:z24}), expresses $\omega(E)$ in terms 
of the regular solutions.   

%
%
\subsection{The well-barrier (WB) model}
We now choose a specific model for the inner potential, consisting of 
an inner square well enclosed by a square barrier. In this case, 
\begin{eqnarray}                       
\phi(k;r) &=& {1\over k_I} \sin k_I r \quad , \quad r<r_a \nonumber \\
\phi(k;r) &=& a_{II} e^{\kappa r} + b_{II} e^{-\kappa r} \quad , 
\quad r\in(r_a,r_d), 
\label{eq:z27}
\end{eqnarray}
where $k_I^2 = k^2 + v_0 $, $v_0 = 2m V_0/\hbar^2$, $\kappa^2 = v_b - 
k^2$ and $v_b = 2m V_b/\hbar^2$. The constants $a_{II}$ and $b_{II}$ 
are determined by matching at $r=r_a$ and from them we get the 
expressions for $\phi(k;r_d)$ and its derivative, 
\begin{eqnarray}                     
\phi(k;r_d) &=& {1\over k_I} \left(\cosh \kappa r_b \sin k_I r_a + 
{k_I \over \kappa} \sinh \kappa r_b \cos k_I r_a \right)\!,
\nonumber \\ 
\phi'(k;r_d) &=& {\kappa \over k_I}\left( \sinh \kappa r_b \sin k_I 
r_a + {k_I\over \kappa} \cosh \kappa r_b \cos k_I r_a \right)\!, 
\nonumber \\ 
&& {\rm  with} \quad r_b=r_d-r_a. 
 \label{eq:z28}
\end{eqnarray}
Inserting these into Eqs. (\ref{eq:z22}) and (\ref{eq:z23}) determines 
the Jost function. The overlap is now given by 
\begin{eqnarray}                      
\la w_E|u_i\ra &= &  {2\over \hbar} \sqrt{{ m k}\over{\pi r_a}}  
{1\over {k_{I} |f(k)|}} \int_0^{r_a} \ dr \sin k_{I} r \sin k_a 
r \nonumber \\ 
 &=&  {2\over \hbar} \sqrt{{ m k}\over{\pi r_a}} \ {{(-)^{n_a}} \over 
{k_{I} |f(k)|}}\ 
 {{k_a \sin k_{I}r_a}\over{(k_a^2-k_{I}^2)}},
\label{eq:z29}
\end{eqnarray}
and therefore when $k_a=\pi/r_a$ ($n_a=1$), 
\begin{equation}                       
\omega(E) = \frac{4\pi\, m k \, \sin^2 k_I r_a}{\hbar^2 r_a^3 
k_I^2 |f(k)|^2 [(\pi/r_a)^2-k_I^2]^2}. 
 \label{eq:z30}
\end{equation}
Since $k_I^2= k^2+v_0$,  even in this simple model the 
$k$-dependence is non-trivial. 
\begin{figure} [htb]                         
\includegraphics[width=8cm]{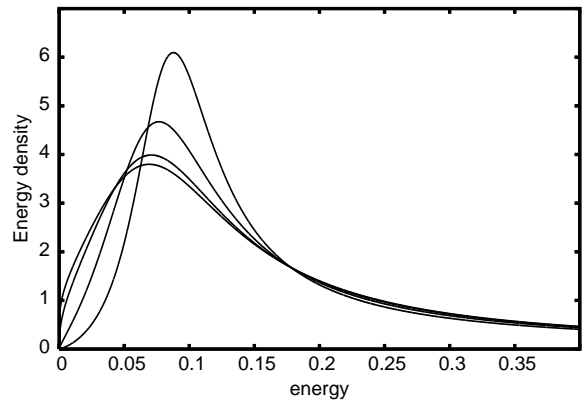}
\caption{Energy densities $\omega(E)$ v.s. energy for $\beta = -0.4$, 
$-0.1, 0.3$ and $0.7$, showing that peak heights increase with $\beta$
but $\omega(E)$ becomes singular for $\beta < 0$.} 
 \label{fidr1}
\end{figure} 
\begin{figure} [htb]                         
\includegraphics[width=8cm]{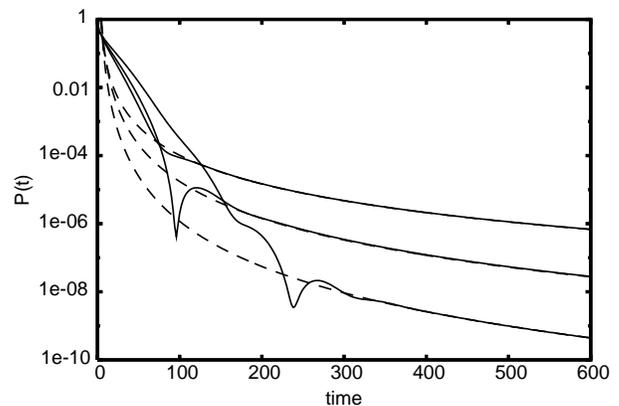}
\caption{Survival probabilities, $P(t)$, exact (solid lines) and 
asymptotic approximation (dashed lines) of  Eq. (\ref{eq:z40}). From 
top to bottom: $\beta = -0.1, 0.3$ and $ 0.7.$ } 
 \label{fidr2}
\end{figure} 
To illustrate our results, we fix on specific parameter values for our inner 
potential. We use units such that $\hbar^2= 2m = 1$ and set 
$r_a= 3.0$, $r_d= 3.4$, $v_0= 0.5$ 
and $v_b = 1.8$.  Fig. \ref{fidr1} shows the computed energy spectra 
when the outer part of the potential has $\beta$ ranging from $-0.4$ 
to $0.7$. Fig. \ref{fidr2} shows the corresponding survival 
probabilities. It can be seen that for this choice of inner 
parameters a strong resonance peak dominates the energy density in 
all cases, and that there is exponential decay of $P(t)$ at 
intermediate times. Other smaller peaks present in $\omega(E)$ at 
higher energies, not shown in the figure, affect 
the corresponding  $P(t)$ only at very small times. We are not  
interested in that regime here. At long times the survival 
probability shows  algebraic decay in all cases. To determine the 
form of $P(t)$ when $t \to \infty$, we need the asymptotic properties 
of $\omega(E)$ near threshold. We find them now in the 
context of the boundary condition model. 
%
\subsection{The threshold energy density} 

From Eq.  (\ref{eq:z18}), 
\begin{equation}                      
w_E(r) \simeq  \sqrt{{2 m k}\over{\pi \hbar^2}} \ \frac{1}
{|f(k)|} \phi_0(r). 
 \label{eq:z31}
\end{equation}
and therefore 
\begin{eqnarray}                      
\la w_E|u_i\ra &\simeq&  {2\over \hbar} \sqrt{{ m k}\over{\pi r_a}} 
{1 \over{|f(k)|}} \int_0^{r_a} \phi_0(r) \sin k_a r \ dr. 
 \label{eq:z32}
\end{eqnarray}
Clearly in this limit all the $k$-dependence is in 
the factor multiplying the integral, while the latter is to this order 
a constant. For the  specific case of the WB model, the analytic 
expression for the integral can be obtained from Eq. (\ref{eq:z29}),  
replacing $k_I$ by $k_{I,0}= \sqrt{v_0}$. 

We look first at the behaviour of the Jost function, Eqs. 
(\ref{eq:z22}) and (\ref{eq:z23}), also to lowest order in $k$. Using 
standard expansions, (see Appendix A,) 
the behaviour when $\beta>0$ as $x= k r_d \to 0$ is 
\begin{eqnarray}                       
\hat{j}_{\beta}(x) &\simeq& {{\pi^{1/2} x^{\beta+1}}\over{2^{\beta+1} 
(\beta+1/2)!}}, 
 \nonumber \\
\hat{n}_{\beta}(x) &\simeq& - {{(\beta-1/2)! \ 
2^{\beta}}\over{\pi^{1/2} x^{\beta}}}, 
 \label{eq:z33}
\end{eqnarray}
and the contribution from the Neumann function will 
dominate in Eq. (\ref{eq:z23}). One easily finds that 
\begin{eqnarray}                      
|f(k)| &\simeq&   {\cal D} \ k^{-\beta}  \quad {\rm with}    \nonumber \\
{\cal D} &\equiv& {{2^{\beta} (\beta-1/2)!}\over {\pi^{1/2} r_d^{\beta} }} \ 
\left|\beta {{\phi_0}\over r_d} + \phi'_0 \right|, 
 \label{eq:z34}
\end{eqnarray}
where $\phi_0\equiv\phi_0(r_d)$ and $\phi'_0 \equiv \phi'_0(r_d)$. 
For the WB model, these are given explicitly in Eqs. (\ref{eq:z28}). \\
Inserting Eq. (\ref{eq:z34}) into Eq. (\ref{eq:z29}) 
we arrive finally at the desired result,  
\begin{equation}                     
\omega(E) \simeq {{4 m \pi  \ \sin^2 (k_{I,0} r_a)}\over { \hbar^2 
r_a^3\ (k_a^2-k_{I,0}^2)^2 \  k_{I,0}^2 \  {\cal D}^2}}\ k^{2\beta+1} 
\equiv \zeta \ k^{2\beta+1}. 
\label{eq:z35}
\end{equation}

\subsection{Decay at long times}
Usually, when an explicit expression for the energy density is 
available, to find analytic approximations for the asymptotic 
contribution to $P(t)$ one replaces the integration path in Eq. 
\ref{eq:z5} by an equivalent one in the complex plane. Examples of 
these methods can be found in \cite{Kn77}, \cite{FGR78} and 
\cite{OK92}. In our case, we choose a closed path consisting of the 
positive real energy axis, a quarter circle of infinite radius 
joining the positive real axis to the negative imaginary axis, and 
the latter. As shown in the Appendix B, integration along the 
circular arc gives a vanishing contribution. Therefore 
\begin{eqnarray}                       
A(t) &=& \int_{0}^{\infty}  dE  \ \omega(E) \ e^{-iEt/\hbar} \nonumber \\
     &=& \oint dE \ \omega(E)\ e^{-iEt/\hbar} + \int_{0}^{-i \infty} 
dE \ \omega(E)  \ e^{-iEt/\hbar} \nonumber \\ 
 &\equiv& A_p(t) + A_v(t).
\label{eq:z36}
\end{eqnarray}
We have analytically continued $\omega(E)$ into the lower half plane, 
and placed the cut running from the origin in the upper half plane. 
The term $A_p(t)$ has contributions from the poles enclosed in the 
contour. These are related to the resonances and lead to 
exponentially decaying terms, so we will not discuss $A_p(t)$ 
further. 

The second term, $A_v(t)$, gives the dominant contribution to the 
decay at long time. To evaluate it, we make the change of variable 
$E=-ix$,  
\begin{equation}                       
A_v(t) = i \int_0^{\infty} \ dx \ e^{-xt} \ \omega(-ix). 
\label{eq:z37}
\end{equation}
In our examples the exponential decay becomes negligible beyond $t 
\simeq 200$, so (assuming $\omega(-ix)$ is smooth), the range of 
values of $x$  giving significant contributions to the integral is  
$0 < x   < x_c \simeq 1/200$. Therefore, in a first approximation 
we use the small $k$ expansion of $\omega(E)$, Eq. (\ref{eq:z35}), to 
extend this function to the relevant part of the negative imaginary 
axis, and write 
\begin{equation}                       
\omega(-ix) \simeq (-i)^{\beta + 1/2}  \zeta  \ x^{\beta+ 1/2}.
\label{eq:z38}
\end{equation}
Inserting this into Eq. (\ref{eq:z37}) one immediately finds 
\begin{equation}                       
A_v(t) \simeq - (-i)^{\beta + 3/2} \zeta \ \Gamma(\beta + 3/2)\ 
t^{-(\beta+ 3/2)}
\label{eq:z39}
\end{equation}
and therefore asymptotically 
\begin{equation}                       
P(t) \simeq |A_v(t)|^2 =  \zeta^2\, \Gamma^2( \beta+ 3/2)\ t^{-(2\beta + 3)}. 
 \label{eq:z40}
\end{equation}
%
\begin{figure} [htb]                         
\includegraphics[width=8cm]{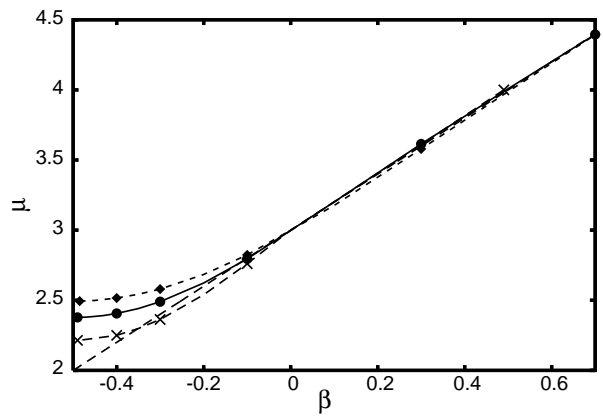}
\caption{ Effective $\mu_f$ v.s. $\beta$. Continuous line with filled 
circles:  standard choice of parameters for the inner part of the 
potential. Dotted  line with filled diamonds: standard set except 
$v_b = 1.6$, dashed line with crosses: standard set except $r_d = 
3.5$. The straight dashed line is $\mu =2\beta+3$. } 
 \label{fidr4}
\end{figure} 

This is the result promised in the Introduction: it shows 
that in this approximation the algebraic exponent is indeed $\mu = 
2\beta + 3$. Fig. \ref{fidr2} confirms graphically the accuracy  of 
this result for a range of values of $\beta$ either repulsive or 
weakly attractive. Note that this result is remarkable in two 
respects: because a) no other parameter of the potential 
affects this prediction; and b) the contribution of the higher 
order terms neglected in writing Eq. (\ref{eq:z33}) is negligible. 
Within the WB model we have checked that other choices of parameters 
for the inner part of the potential lead to the same conclusions. 
This will be further confirmed in figure \ref{fidr4}. 

\begin{figure} [htb]                         
\includegraphics[width=8cm]{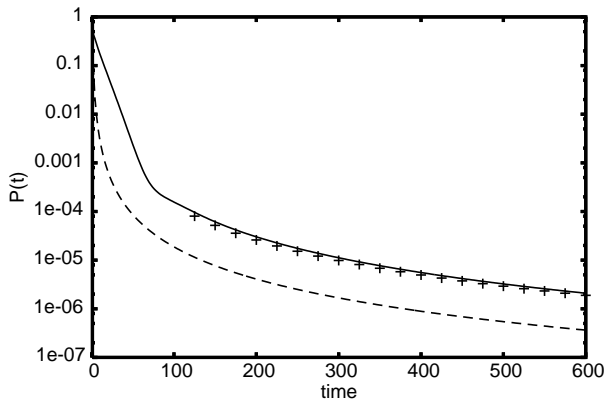}
\caption{ Survival probability, $P(t)$, for $\beta = -0.4$, 
($\nu = 0.1$). Continuous line: exact; dashed line: single term 
asymptotic approximation Eq. (\ref{eq:z40}); crosses: 
the expansion of Eq. (\ref{eq:z45}) truncated at 4 terms.} 
 \label{fidr3}
\end{figure}

However, the above approximation fails  for values of $\beta$ that 
make the outer potential attractive, especially when approaching 
the lowest value we considered, $\beta = -1/2$. This is shown in 
Fig. \ref{fidr3} for $\beta=-0.4$. Empirically we have found that for 
the range of times shown in the figure the asymptotic decay can still 
be well fitted by an algebraic form, but the exponent determined from 
such a fit deviates from the simple law that holds for positive 
$\beta$. Intuitively, the simple 
algebraic behaviour of the repulsive case $\beta>0$ may be understood 
as a consequence of the dominance of the long centrifugal-like tail 
of the potential in near threshold scattering. However if the tail is 
atractive the particle is drawn to explore the inner part of the 
potential. This makes the decay more complex and also more sensitive 
to this inner part.    
We will now discuss that in further detail.   

Looking again at the expansions of the Ricatti-Bessel functions in 
Eq. (\ref{eq:za2}) one sees that when $\beta \to -1/2$ the series for 
$\hat{j}_{\beta}$ and $\hat{n}_{\beta}$ both begin with 
$x^{1/2}$ and therefore the contribution 
from the  $\hat{j}_{\beta}$ cannot be discarded. 
At this stage it is convenient to use the auxiliary index, $\nu=\beta 
+1/2$, that corresponds to the index of the cylindrical Neumann and 
Bessel functions in Eq. (\ref{eq:za1}). We will be interested in 
small values of $\nu$. Our representative example will be that of 
Fig. \ref{fidr3}, $\nu=0.1$. Including only the lowest order terms, 
$p=0$ in Eq. (\ref{eq:za2}), but for both functions, one 
finds 
\begin{equation}                       
|f(k)|^2 \simeq  k \left(\lambda_- k^{-2\nu} + \lambda_0 + 
\lambda_+ k^{2\nu} \right) \ 
\label{eq:z41}
\end{equation}
with
\begin{eqnarray}                       
\lambda_- &\equiv& {{[\Gamma(\nu)]^2 \ 2^{2\nu-1}}\over{\pi 
r_d^{2\nu-1}}} \left[\left(\nu-{1\over 2}\right){\phi_0\over r_d} 
+\phi'_0\right]^2
\nonumber \\
\lambda_0 &\equiv& {{\cot( \nu \pi) \ r_d}\over \nu} \left[\left(\nu-
{1\over 2}\right){\phi_0\over r_d} +\phi'_0\right] \times
\nonumber \\ 
&& \qquad \left[\left(\nu+{1\over 2}\right){\phi_0\over r_d} -\phi'_0\right]
\nonumber \\
\lambda_+ &\equiv& {{r_d^{2\nu+1} [\Gamma(1-\nu)]^2} \over {\pi 2^{2\nu+1} 
\nu^2}} \left[\left(\nu+{1\over 2}\right){\phi_0\over r_d} -\phi'_0\right]^2. 
\nonumber \\ 
 \label{eq:z42}
\end{eqnarray}
Therefore 
\begin{equation}                       
\omega(E) \simeq {{\zeta \ k^{2\nu}}\over{1+ (\lambda_0/\lambda_-) 
k^{2\nu} + (\lambda_+/\lambda_-) k^{4\nu}}} 
\label{eq:z43}
\end{equation}
replaces Eq. (\ref{eq:z35}). To have a result for $A_v(t)$ 
similar in form to the earlier one, we expand the denominator and write 
\begin{equation}                       
\omega(E) \simeq \sum_{m=1}^{\infty} \zeta_{2m}\, k^{2m \nu}, 
\label{eq:z44}
\end{equation}
with $\zeta_2 = \zeta$ already defined, and the other $\zeta_{2m}$ 
obtained from Eqs. (\ref{eq:z42}) and (\ref{eq:z43}). Repeating 
the derivation in Eqs. (\ref{eq:z38}) to (\ref{eq:z40}), we obtain 
\begin{equation}                       
A_v(t) \simeq - \sum_{m=1}^{\infty} \ \zeta_{2m}\ 
\Gamma(1 + m\nu)  \ (it)^{-(1+ m \nu)}.  
\label{eq:z45}
\end{equation}
The first term in this expansion reproduces Eq. (\ref{eq:z39}), but 
when $\nu \simeq 0.1$ the algebraic exponents of the first 
few terms in the series will be comparable, and keeping only the 
first would be insufficient.  Confirming this, figure \ref{fidr3} shows 
that with a single term in the series the prediction is wrong by an 
order of magnitude, and that one has to include at least four terms 
in the series to reproduce the exact $P(t)$ for $t > 200$. When we 
use the approximation for $\omega(E)$ of Eq. (\ref{eq:z43}) in Eq. 
(\ref{eq:z37}), the agreement is even better. 

These results show that  for positive $\beta$ one can neglect the 
$k$-dependence in $\phi(k,r_d)$ and its derivative, and also the 
higher order terms in the series for the Ricatti-Bessel functions, 
and still have a satisfactory approximation for the asymptotic decay 
of the survival probability over the range of time included in the 
figures. But when $\beta$ becomes negative (outer potential 
attractive) the truncation leading to the asymptotic law in 
Eq. (\ref{eq:z40}) ceases to be valid. Even then, we have empirically 
found that the exact $P(t)$ can be well fitted  with a formally 
similar algebraic expression in a range of times that we will specify 
as the interval $400 < t < 800$. To be more precise, and to try to 
simulate the way in which the algebraic exponents could be extracted 
in experiments like that of Rothe {\it et al},  we have made linear 
least square fits to  the exact values of $\ln P(t)$  v.s. $\ln t$ 
for equally spaced values of $t$ in the above range. The fits are 
always excellent, so that in that range of times one can hardly 
distinguish in the figures the exact $P(t)$ from the fitted form 
$P_f(t) = {\cal M}\ t^{-\mu_f}$. The effective algebraic exponents, 
$\mu_f$, thus found are shown in figure \ref{fidr4}. One sees that 
when $\beta <0$ the effective exponents depend on the parameters of 
the potential, and deviate from the simple law  $\mu=2 \beta+3$. 
Making the barrier wider or higher, increases the values of $\mu_f$ 
for a given $\beta$, but in all cases as $\beta \to -1/2$ one finds 
that the values of $\mu_f$ become constant.

We tried to derive explicit analytic expressions for $P(t)$ 
in the limit $\nu=0$. Using only the first order term in the 
expansion of the Neumann function, $N_0(z)$, the first 
term in the asymptotic series for $A_v(t)$ goes like $1/[t 
\ln(t/r_d^2)]$ with the next order terms involving higher 
powers of the logarithm. However, using only the first order term of 
the series for $N_0$ is not enough; one needs at least one more 
term in both $J_0$ and $N_0$. With these added terms, we were 
unable to find a useful analytic expression for $A_v(t)$. Still, the 
fact that the lowest order results include $\ln t$ and powers thereof 
indicates that the algebraic form of decay for the asymptotic 
part of $P(t)$ is not as universal as previously thought. 
It is only for potentials with a vanishing or repulsive 
outer part that such a simple algebraic decay law can be 
established.

\section{Discussion} 
Our study of the long-time deviations from exponential decay, was 
motivated by the apparent contradiction between the recent (and so 
far only) experimental results which unambiguously show such 
deviations, and theoretical models: the power laws in the experiments 
have non-integer exponents whereas the models predict or postulate 
integers. What is the physical origin of non-integral exponents? The 
complexity of the system studied experimentally (large, excited 
organic molecules in solution), made us consider, instead of an {\it 
ab initio} or realistic approach, the more modest goal of 
understanding and answering the question in a tractable system. 

Limiting ourselves to a single-particle model of wave-packet decay 
from a scattering potential region,  we have established a link 
between non-integer exponents and long potential tails, which is 
expressed by a simple formula for repulsive inverse square 
potentials. Before arriving at these results we explored several 
other possible sources of ``anomalous'' long-time decay: in 
particular, we tried to simulate the effects of the solvent molecules 
and of temperature by introducing  randomly fluctuating perturbations 
in the potential and averaging over many realizations.   This lead to 
localization effects and a non-vanishing constant value of $P(t)$ at 
long times but non-integral exponents were not found. Another attempt 
was consideration of complex potentials that could simulate the 
effect of coupled channels due to measurement or other effects. 
However   the result of an imaginary term is similar to that 
described in \cite{DMG06}, Eq. (11): the long-time deviation is altered 
but not in the sought for algebraic form.

Despite our retreat from the arena of the organic molecule 
experiments, the long-time decay described here is not simply 
academic. Attractive inverse square potentials occur physically as 
effective radial potentials between a charged wire and a polarizable 
neutral atom \cite{S98}, the strength factor being proportional to 
the square of the linear charge density of the wire and thus 
controllable \cite{S98,S99,Gol92}. Combined with a repulsive 
centrifugal term, an arbitrary $\alpha/r^2$ potential may be 
implemented. In addition, it is possible to modify the inner region 
and implement a potential minimum by a time varying sinusoidal 
voltage in the high frequency limit \cite{Gol92}, or by replacing the 
wire by a charged optical fiber with blue detuned light propagating 
along the fiber and the cladding removed \cite{S99}.   Decay 
experiments with cold atoms showing exponential laws have been 
performed \cite{S98}, and the ability to modify the potential 
parameters makes the observation and study of the long-time power-law 
in these systems a realistic prospect.

\section{Acknowledgements}

We are grateful to DGES-Spain for support through grants  
FIS2004-03156, FIS2006-10268-C03-01;   
to UPV-EHU (grant 00039.310-15968/2004); and to  NSERC-Canada for 
Discovery Grant RGPIN-3198 (DWLS).

\appendix 
\section{}
The Riccati-Bessel functions $\hat{j}_\beta(x), \hat{n}_\beta$,
and $\hat{h}^\pm_\beta(x)=\hat{n}_\beta\pm i\hat{j}_\beta$
are defined as $x$ times the corresponding spherical functions, or, 
in terms of cylindrical Bessel and Neumann functions,  
\begin{eqnarray}                        
\hat{j}_{\beta}(x) &=& \sqrt{{\pi x}\over 2} J_{\beta+1/2}(x),
\nonumber \\
\hat{n}_{\beta}(x) &=& \sqrt{{\pi x}\over 2} N_{\beta+1/2}(x).
\label{eq:za1}
\end{eqnarray}
In this paper $\beta$ may be non-integer.

Their series representations are    
\cite{Grad},  
\begin{eqnarray}                       
\hat{j}_{\beta}(x) &=& \sqrt{\pi} \left({x\over 2}\right)^{\beta+1} 
\sum_{p=0}^{\infty} {{(-)^p}\over{p! \ \Gamma(\beta+p +3/2)}} 
\left({x\over 2}\right)^{2p},
\nonumber \\ 
\hat{n}_{\beta}(x) &=& \cot \left(\beta+{1\over 2}\right)\pi \ \hat{j}_{\beta}(x) 
 - {{\sqrt{\pi}}\over{\sin (\beta+1/2) \pi}}
\nonumber \\ 
&& \times \left({x\over 2}\right)^{-\beta} \sum_{p=0}^{\infty} {{(-
)^p }\over{p!\ \Gamma(p -\beta +1/2)}} \left({x\over 2}\right)^{2p}. 
\label{eq:za2}
\end{eqnarray}

\section{}
We will prove here that the energy density extended to the quarter 
circle at infinity, in the fourth quadrant of the energy plane, is 
vanishingly small and therefore does not contribute to $A(t)$ in Eq. 
(\ref{eq:z36}). To do so, we first rewrite the energy density as
\begin{eqnarray}                    
\omega(E) &=& {{4m \pi}\over{\hbar^2 \pi^3 r_a^3 k_I^2 ((\pi/r_a)^2-
k_I^2)^2}} \ {\cal G}(E)   \nonumber \\
{\cal G}(E) &=& {{\sin^2 k_I r_a}\over {\cal C}^2}. 
\label{eq:zb1}
\end{eqnarray}
We will now extend these functions to the fourth quadrant in the 
energy plane and look at their asymptotic behaviour when $|E| \to 
\infty$. It is convenient to write
\begin{eqnarray}                    
E &=& |E| e^{2 i \varphi},
\nonumber \\
k &=& |k| e^{i\varphi},
\label{eq:zb2}
\end{eqnarray}
and $\varphi \in(-\pi/4,0)$. Similarly we will write $k_I = |k_I| 
e^{i \varphi_I}$ and will now find an approximate expression for the 
latter valid in the limit when $|k| \to \infty$. From $k_I^2 = k^2 + 
v_0$, we can write
 \begin{eqnarray}                    
|k_I|^2 e^{2i\varphi_I} &=& |k|^2 e^{2i\varphi} + v_0 \nonumber \\
 &\simeq& \left( |k|e^{i\varphi} + {v_0\over{2|k| 
e^{i\varphi}}}\right)^2,
\nonumber \\ 
|k_I| e^{i\varphi_I} &\simeq& |k|e^{i\varphi} + {v_0\over{2|k|}} e^{-
i\varphi} + O\left(|k|^{-2}\right),
\nonumber \\ 
k_I &\simeq& k + {v_0\over{2|k|}} e^{-i\varphi} + O\left(|k|^{-
2}\right). 
 \label{eq:zb3}
\end{eqnarray}
Similarly, since $k_b^2= k^2- v_b$, 
\begin{equation}                     
k_b \simeq k - {v_b\over{2|k|}} e^{-i\varphi} + O\left(|k|^{-2}\right).
\label{eq:zb4}
\end{equation}
In the same limit, the sine in the numerator of ${\cal G}(E)$ will give
\begin{equation}                     
\sin k_I r_a \simeq {1\over {2i}} e^{ik_I r_a} \to \infty 
\label{eq:zb5}
\end{equation}
since Im$(k_I)<0$. Let us now consider the denominator: from 
Eq. (\ref{eq:z23}) one sees that both the Ricatti-Bessel functions and 
the regular solutions at $r=r_d$ must be extended to complex energies 
in the fourth quadrant and in the large $|E|$ limit. Using the 
asymptotic expansions given e.g. in \cite{Ar85}, 
we 
find 
\begin{eqnarray}                    
{\hat n}_{\beta}^2+ {\hat j}_{\beta}^2 &\simeq &  1+  \frac{\beta 
(\beta+1)}{2z^2}   + O(z^{-4}) \nonumber \\ 
 {\hat n}_{\beta} {\hat n}'_{\beta} + {\hat j}_{\beta}{\hat 
j}'_{\beta} &\simeq& -\frac{\beta  (\beta+1)}{2 z^3} + O( z^{-5}) \nonumber \\ 
 ({\hat n}'_{\beta})^2 + ({\hat j}'_{\beta})^2 &\simeq&  
1- \frac{\beta(\beta+1)}{2z^2} + O(z^{-4}), 
 \label{eq:zb6}
\end{eqnarray}
with $z=k r_d$. Inserting these into Eq. (\ref{eq:z23}) we find to 
order $z^{-4}$
\begin{eqnarray}                    
{\cal C}^2  &\simeq& \phi^2+ {1\over k^2} (\phi')^2 + 
\frac{\beta(\beta+1)}{2z^2} \left(-\phi^2+ {1\over k^2} (\phi')^2 \right) 
\nonumber \\ 
&+& 2 \phi \phi' {1\over k} \ \frac{\beta(\beta+1)}{2z^3}. 
\label{eq:zb7}
\end{eqnarray}
Now one has to deal with $\phi$ and $\phi'$, whose explicit 
expressions are given in Eqs. \ref{eq:z28}. Using the approximations 
in Eqs. (\ref{eq:zb3}, \ref{eq:zb4}) and (\ref{eq:zb5}), one finds 
\begin{eqnarray}                    
\phi^2+{{ (\phi')^2}\over k^2} &\simeq&   -  {{v_b+v_0}\over {4k^4}} 
\ e^{2i(k_b r_b+ k_I r_a)}
\nonumber \\ 
{1\over k^2} \left(-\phi^2+{1\over k^2} (\phi')^2\right) &\simeq&  
{8\over{k^4}}  e^{2i(k_b r_b+ k_I r_a)}
\nonumber \\ 
{1\over k^4} \phi \phi' &\simeq& -{1\over {4k^5}}  e^{2i(k_b r_b+ k_I 
r_a)}, 
 \label{eq:zb8}
\end{eqnarray}
so that all terms in the denominator of ${\cal G}(E)$ diverge like 
$\exp(2i (k_I r_a + k_b r_b))$. Since the $\sin^2$ in the numerator 
diverges only like exp$(2i k_I r_a)$, see Eq. (\ref{eq:zb5}), the result 
is that $|{\cal G}(E)|  \to 0$ and therefore the contribution to $A(t)$ 
due to the integral on the quarter circle at infinity is nil. 

Note that this result can be easily extended to any value of $n_a$ 
introduced in Eq. (\ref{eq:z25}). Therefore any initial state, 
$|u_i\ra$,  that can be expanded as a finite superposition of 
eigenstates of the infinite square well will have an energy density 
with similar properties.


\end{document}